\documentclass[review]{elsarticle}

\usepackage{lineno,hyperref}
\modulolinenumbers[5]

\journal{Journal of \LaTeX\ Templates}

\begin{document}

\begin{frontmatter}

\title{Initial performance of the high sensitivity alpha particle detector at the Yangyang underground laboratory}

\author[a]{C.~Ha}
\cortext[mycorrespondingauthor]{Corresponding author}
\ead{changhyon.ha@gmail.com}

\author[b]{G.~Adhikari}
\author[b]{P.~Adhikari}

\author[a]{E.~J.~Jeon}
\author[a]{W.~G.~Kang}
\author[a]{B.~H.~Kim}
\author[a]{H.~Kim}

\author[a,b]{Y.~D.~Kim}
\author[a,c]{Y.~H.~Kim}

\author[a]{H.~S.~Lee}
\author[a]{J.~H.~Lee}
\author[a]{M.~H.~Lee}
\author[a]{D.~S.~Leonard}
\author[a]{S.~L.~Olsen}

\author[a,1]{J.~S.~Park}
\fntext[1]{present address : High Energy Accelerator Research Organization (KEK), Ibaraki 319-1106, Japan}

\author[a]{S.~H.~Yong}
\author[a]{Y.~S.~Yoon}

\address[a]{Center for Underground Physics, Institute for Basic Science (IBS),\\Daejeon 34047, Republic of Korea}
\address[b]{Department of Physics, Sejong University,\\Seoul 05006, Republic of Korea}
\address[c]{Korea Research Institute of Standards and Science,\\Daejeon 34113, Republic of Korea}

\begin{abstract}
  Initial performance of the UltraLo-1800 alpha particle detector
  at the 700~m deep Yangyang underground laboratory in Korea
  is described. The ionization detector uses Argon as a counting gas
  for measuring alpha events of a sample. 
  We present initial calibration results and low-activity sample measurements
  based on the detector's pulse discrimination method and a hardware veto.
  A likelihood analysis that shows a separation of a bulk component from a surface component with a contamination depth
  from $^{210}$Po alpha particles using simulated models is presented.
\end{abstract}


\end{frontmatter}

\linenumbers

\section{Introduction}
Rare event searches including dark matter~\cite{Undagoitia:2015gya} and neutrinoless double beta decay~\cite{Ostrovskiy:2016uyx} experiments require detector materials and environments
with the lowest possible levels of radioactive contamination
because the level of contamination often directly limits their experimental sensitivities.
Efforts towards a selection of low-radioactivity materials, development of cleaning techniques,
and environmental control are considered as a standard procedure in these experiments~\cite{Akerib:2017iwt,Aprile:2017ilq,Henning:2016fad,Leonard:2017okt,Zuzel:2018lrt}.

\begin{figure}[!htb]
  \begin{center}
    \begin{tabular}{cc}
      \includegraphics[width=0.75\textwidth]{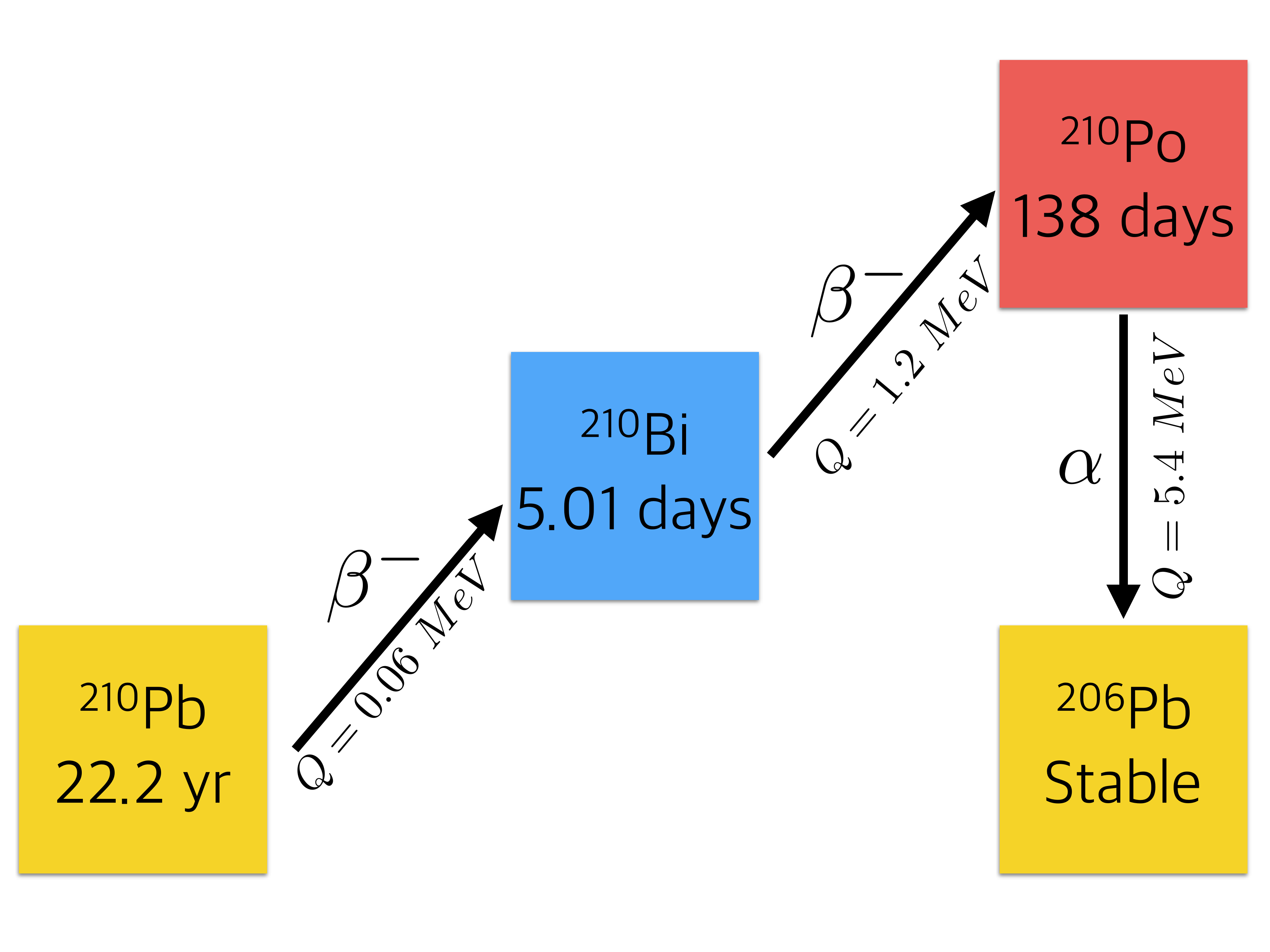}
    \end{tabular}
  \end{center}
  \caption{The $^{210}$Pb decay scheme. $^{210}$Pb decays with a half-life of 22.2 years through $^{210}$Bi to $^{210}$Po and
    then with a half-life of 138 days to $^{206}$Pb. The $^{210}$Pb contamination on a detector material,
    therefore, could be critical because it can persistently affect
    backgrounds at low energies via the beta decays as well as at the high energies through the alpha decays.
  }
  \label{pb210diagram}
\end{figure}

The most common techniques for measuring activity of daughter nuclides of the naturally occurring Uranium and Thorium decays typically measure the Uranium and Thorium concentration of the bulk material directly.
Otherwise they are sensitive only to highly penetrating radiation, particularly gamma emissions which provide at best weak sensitivity to the spatial distribution of the source and thus effectively also sample the bulk of the material. Furthermore, these methods are only directly sensitive to a subset of the daughter decays within the decay chains.
However short range radiation, particularly alpha and beta emissions, is generally generated by different combinations of daughter nuclei within the chains, and only the portion on or near the inner surfaces of a rare-event detector can penetrate to the active detection region and produce backgrounds.  For these reasons it is necessary to directly measure these short range surface emissions using dedicated techniques. $^{210}$Pb, being long lived\,(t$_{1/2}$=22.2\,years), can often plate out onto surfaces (or into the bulk during production processes) exposed to $^{222}$Rn decays in the air, and can thus exist at levels strongly out of equilibrium with the rest of the decay chain.
As a consequence, $^{210}$Pb decays to $^{210}$Bi by emitting a beta particle which can be a background component at the low energies below 60\,keV while $^{210}$Po decays to $^{206}$Pb by producing a 5.3\,MeV alpha particle and a $^{206}$Pb nuclear recoil which may affect region of interest of the rare decay experiments (see Fig.~\ref{pb210diagram}).

Alpha spectroscopy, in particular, allows us to access to these decay chains directly by measuring energy and peak-shape information.
In particular, energy information allows statistical separation of contaminants on the surface and those originating near the surface from sources distributed within the bulk. This helps to understand how to control the background source and how to model its effects on experimental backgrounds. 

\section{Alpha counter}
\begin{figure}[!htb]
  \begin{center}
    \begin{tabular}{cc}
      \includegraphics[width=0.75\textwidth]{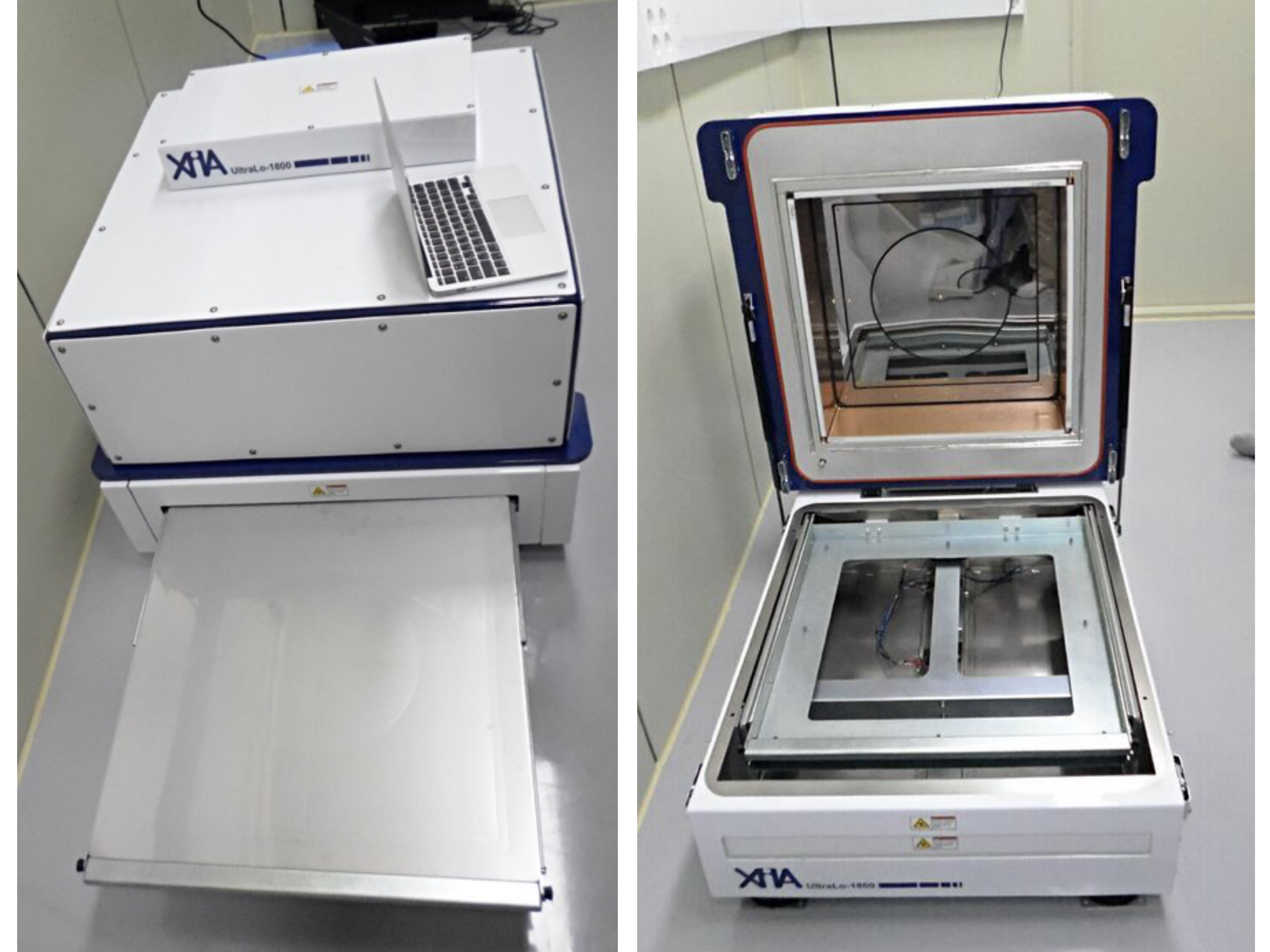}
    \end{tabular}
  \end{center}
  \caption{Two photos of the UltraLo-1800 counter at Y2L.
    The left photo shows the slide-open tray while the right photo shows the opened chamber.
    The clean room that hosts the alpha counter is air-controlled area where low Radon in air,
    low humidity, and stable temperature are provided.
  }
  \label{alphacounter}
\end{figure}

Alpha particle counting using an ionization chamber is a well-known method to estimate a material's radioactivities~\cite{knoll}.
The UltraLo-1800 alpha counter~\cite{ultraloPatent,ultraloIEEE,ultraloBkg,manual} was installed at the A5 tunnel of
the Yangyang Underground Laboratory (Y2L) in June, 2015 (see Fig.~\ref{alphacounter}).

The main purpose for this counter is to
understand the surface contamination of a sample by measuring those alpha events directly from $^{210}$Po decays.
In particular, the alpha detector has served to assay detector materials for the COSINE dark matter
experiment~\cite{Adhikari:2017esn} and the AMoRE double beta decay experiment~\cite{Alenkov:2015dic}.

This instrument records characteristic signals from ionization electrons
produced by a material's alpha emissions in an Ar-filled gas chamber.
The distinct risetime is used to select alpha particles that originate from
the specimen tray and veto those from other locations.
An additional hardware veto using inner and outer electrodes improves the detector sensitivity
because it tags the incoming background events which are produced at the side part of the chamber.

\begin{figure}[!htb]
  \begin{center}
    \begin{tabular}{cc}
      \includegraphics[width=0.75\textwidth]{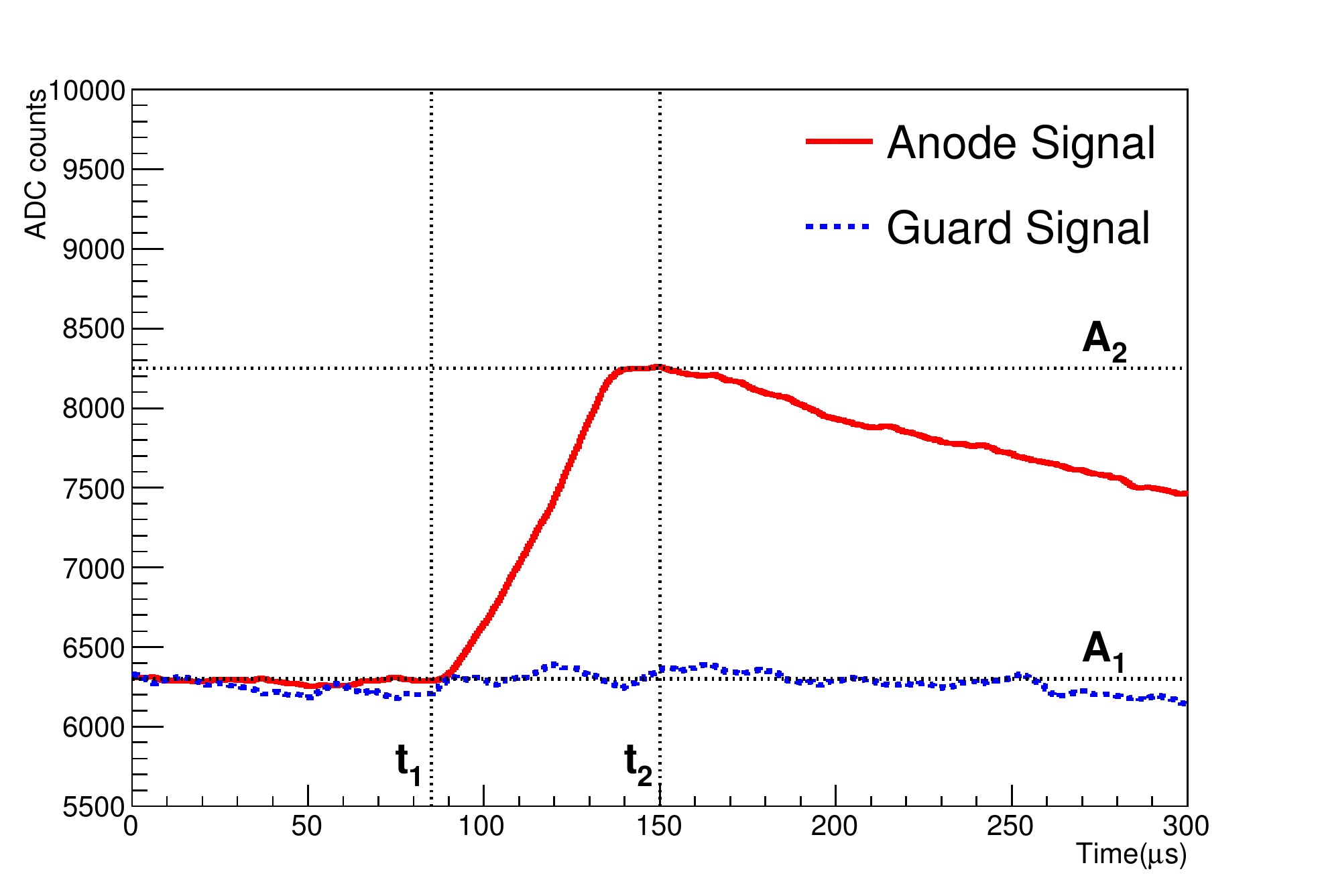}
    \end{tabular}
  \end{center}
  \caption{An example alpha event waveform.
    The event is selected from $^{241}$Am (E$_\alpha$=5.5 MeV) source calibration data.
    The risetime ($t_2 - t_1$) is used in the pulse discrimination method and
    the pulse height ($A_2- A_1$) is converted into energy.
    Note that the $^{241}$Am source hole was covered with a Mylar of 10~$\mu$m thickness and
    so the pulse height has been reduced by about 700 ADC counts.
  }
  \label{waveform}
\end{figure}

The dimension of the ionization chamber is 42$\times$42~cm$^2$ in area and 15~cm in height.
A high voltage of 1100~V is applied between the electrodes and the tray to create a uniform electric field.
When a charged alpha particle ionizes Argon gas molecules in the chamber,
ion pairs (Ar$^+$, e$^-$) are generated along the particle's passage.
According to the Schokley-Ramo theorem~\cite{ultraloShockley,ultraloRamo}, an induced charge on an electrode is linearly related to the potential
difference of the electron's travel distance in an electric field.
Since the generated pulse shape of an alpha signal is determined by the travel time of the induced drift electrons~\cite{ultraloPSD},
the machine identifies alpha particles from the sample area and rejects those produced from other positions of the chamber volume.
Figure~\ref{waveform} shows a typical sample
alpha waveforms in the calibration data.

\begin{figure}[!htb]
  \begin{center}
    \begin{tabular}{cc}
      \includegraphics[width=0.75\textwidth]{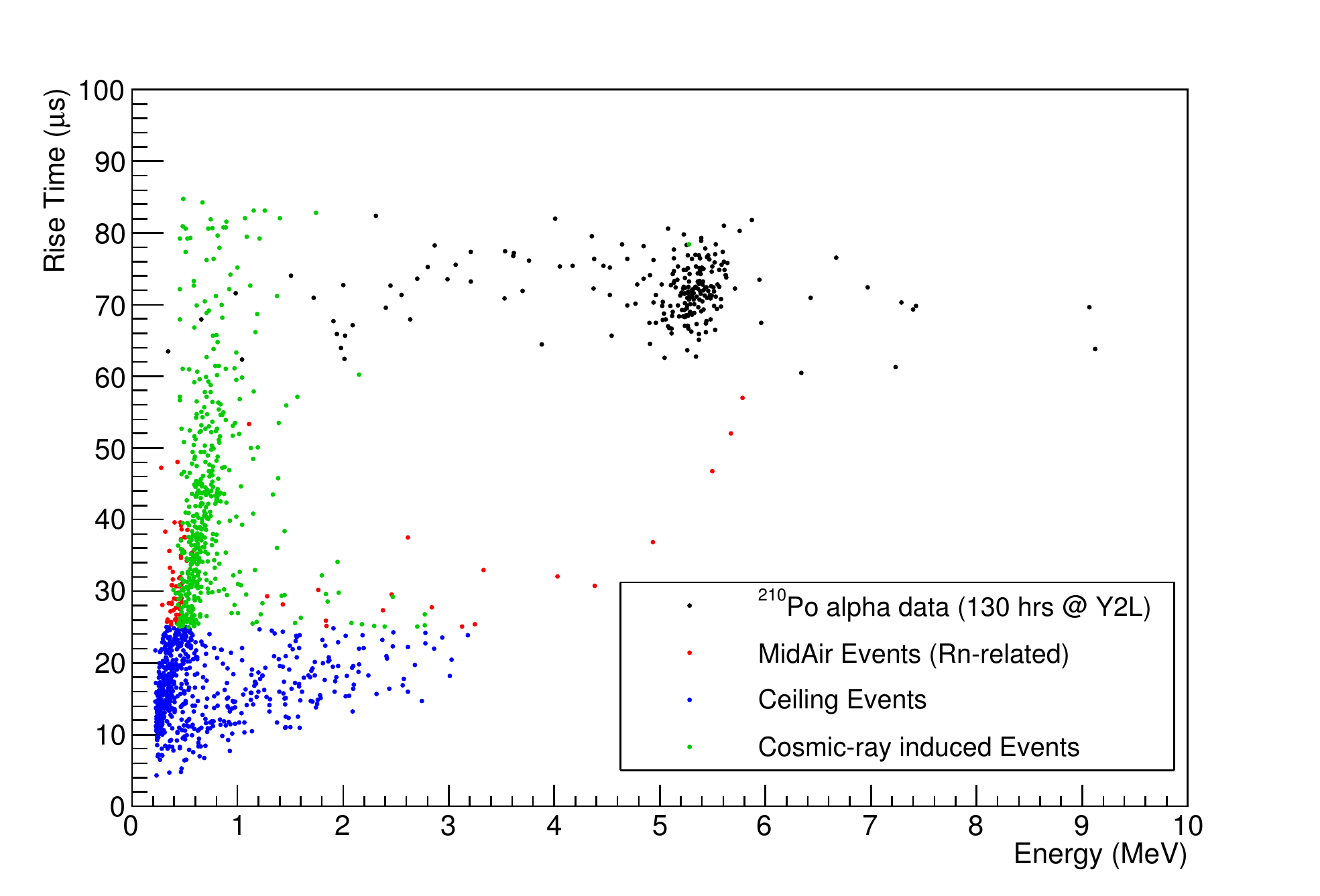}
    \end{tabular}
  \end{center}
  \caption{Pulse discrimination with $^{210}$Po source calibration at Y2L.
    After veto selections (see texts), events above 60~$\mu$s are defined as alpha events (black dots).
    Here vetoed events including cosmogenic, ceiling, and mid-air components are also displayed.
  }
  \label{pid}
\end{figure}

For a measurement, an emissivity ($\epsilon$) is defined as alpha particle counts per hour in a unit area
(cm$^2$). There are two discrimination techniques for this detector; a hardware veto and a software
veto. The hardware veto is applied by a dual-readout configuration with a fiducial (inner) electrode and
a veto (outer) electrode. An initial background suppression is accomplished by rejecting events that pass only
from the veto electrode. The second suppression is performed by using
pulse shape discrimination via the risetime and amplitude of an event (see Fig.~\ref{pid}).

Events with a short risetime are classified as particles originated from a
ceiling. For a $^{210}$Po decay at the sample surface, the risetime of an emitted alpha
was measured to be greater than 60~$\mu$s, and the pulse height is around 2700 ADC counts.
An underground measurement benefits from the fact that the cosmic-ray muon-induced background
events are significantly reduced. The detector shown in Fig.~\ref{alphacounter} is currently
hosted in a room where air quality, humidity, and temperature are controlled.
The detector room temperature is maintained at 23.5$\pm$0.3$^\circ$C, relative humidity is 40$\pm$3\%,
and radon level is measured at around 40\,Bq/m$^3$. 
With a dedicated Ar gas supply, a maximum length of one month measurement without stoppage is
possible in this setup.

\begin{figure}[!htb]
  \begin{center}
    \begin{tabular}{cc}
      \includegraphics[width=0.75\textwidth]{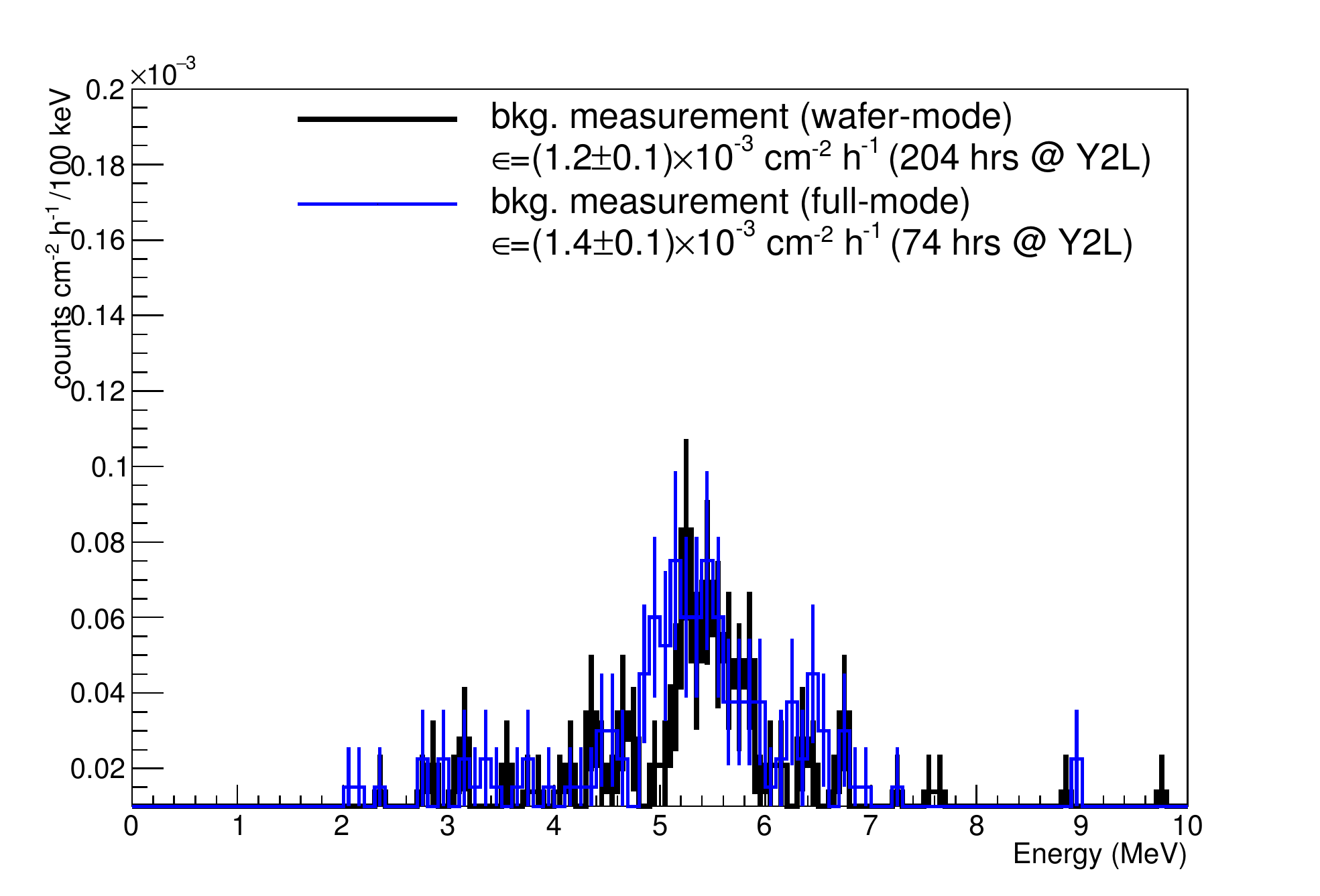}
    \end{tabular}
  \end{center}
  \caption{These plots show two modes of operation in the alpha counter. The backgrounds in both modes show consistent results.
  }
  \label{bkglevel}
\end{figure}

There are two modes of operation: the wafer mode and the full mode. In the full mode, the
sample area becomes 1800~cm$^{2}$ while the wafer mode covers 707~cm$^{2}$.
The area-normalized background levels for the two modes are consistent as shown in Fig.~\ref{bkglevel}.
Samples are prepared with an N$_2$ flux in a glove box and are loaded to the detector tray promptly to prevent additional radon contamination that might
happen while opening the tray. To further remove the residual radon effect, early data (a
few hours) is typically removed at the offline analysis level.

\begin{figure}[!htb]
  \begin{center}
    \begin{tabular}{cc}
      \includegraphics[width=0.75\textwidth]{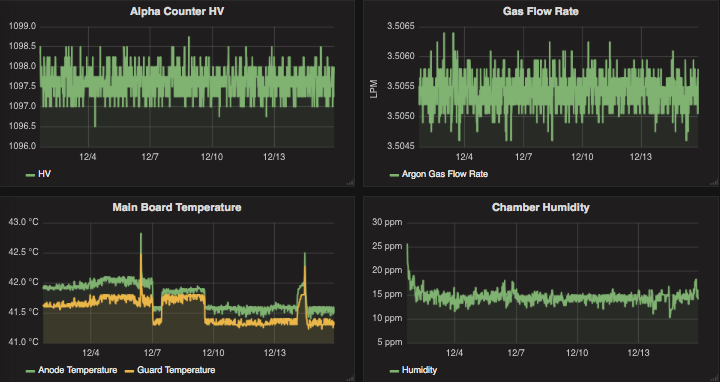}
    \end{tabular}
  \end{center}
  \caption{The alpha counter monitoring system at Y2L. The detector status is regularly monitored by shift takers online and by on-site maintenance staffs. In clockwise, high voltage, gas flow rate, humidity in the chamber, and main board temperatures are shown.
  }
  \label{grafana}
\end{figure}

Several parameters from the alpha counter are monitored in the Y2L's centralized monitoring
system. The long-term monitoring system improves the usage efficiency of the detector and provides
simpler maintenance. Figure~\ref{grafana} shows selected monitoring parameters for the alpha counter.
The counter is remotely controlled and raw data is automatically transferred to the ground office
so that ,other than a sample change, there is no need to be inside the detector room.
\section{Calibration}
\begin{figure*}[!htb]
  \begin{center}
    \begin{tabular}{cc}
      \includegraphics[width=0.45\textwidth]{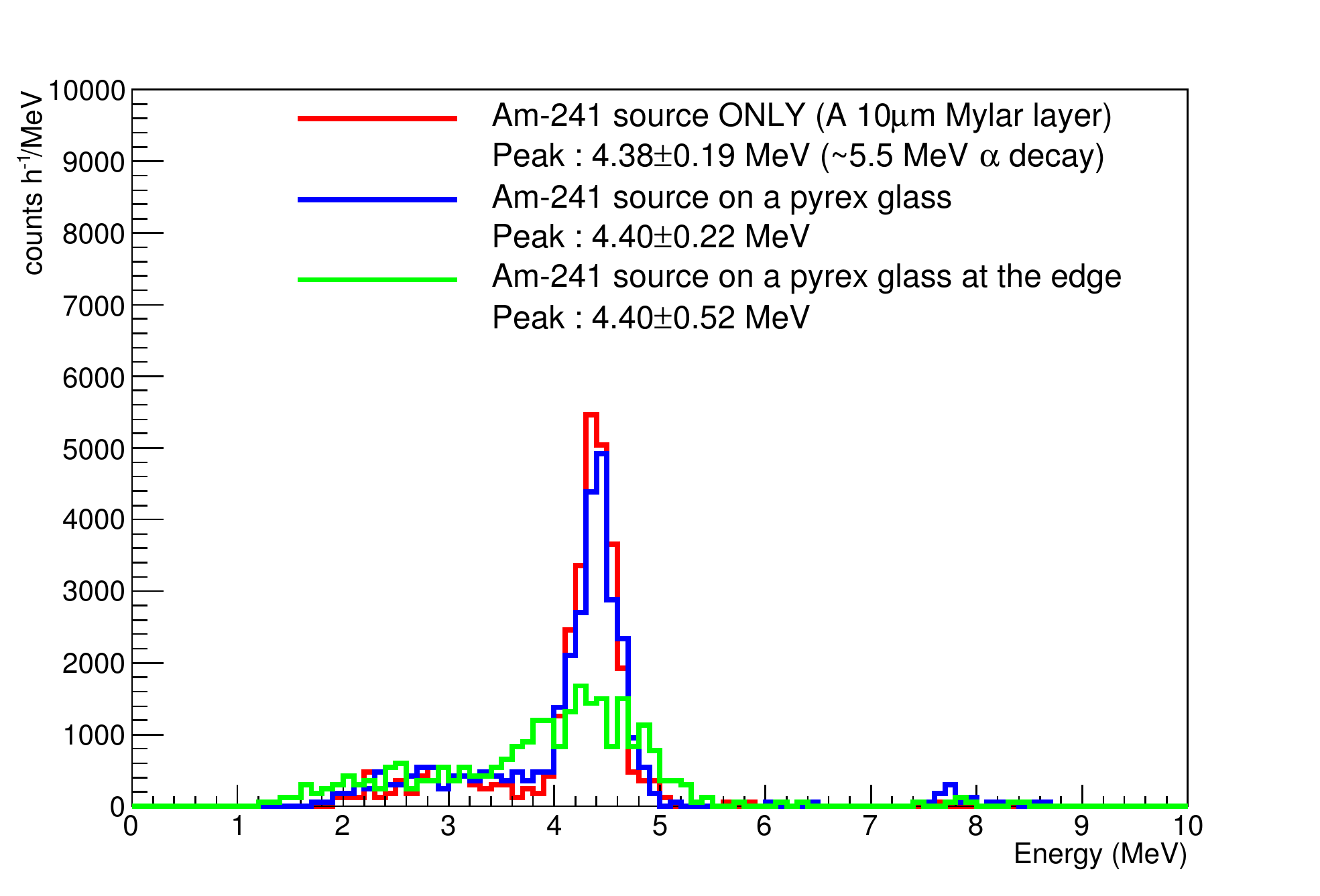} &
      \includegraphics[width=0.45\textwidth]{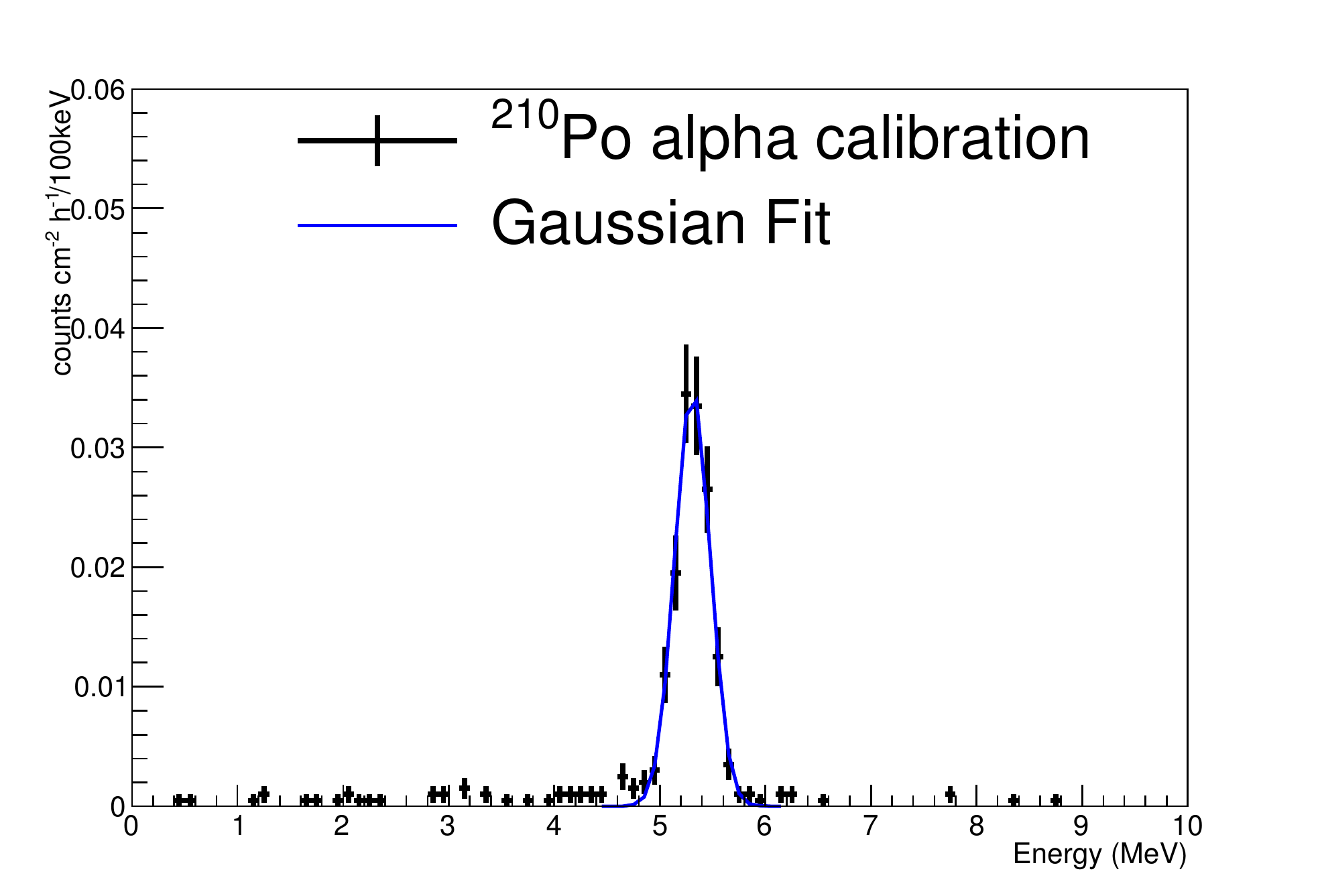} \\
    \end{tabular}
  \end{center}
  \caption{Source calibrations at Y2L using $^{241}$Am and $^{210}$Po radioactive alpha emitters.
  $^{241}$Am is pinhole source while the $^{210}$Po is deposited on a square copper plate.
  Left plot shows position dependence vs. the resolution and stability with a dielectric plate placed under the source
  for the $^{241}$Am source. The degradation of the resolution is seen when the source is placed at the edge of the detection area
  due to the partially escaped track.
  Right plot shows the best resolution achieved with the $^{210}$Po source.
With Gaussian fit sigmas, 4.3\% and 3.2\% resolutions have been achieved
           for the two sources, respectively.
           }
  \label{ULcalibration}
\end{figure*}

Although the detector was calibrated for the energy and risetime at the company prior to delivery,
separate calibrations with different calibration sources were performed regularly
to check the energy scale and its long-term stability.
An $^{241}$Am pinhole source (the 5.5~MeV alpha emitter) and 5~cm$\times$5~cm laminar $^{210}$Po source (the 5.3~MeV alpha emitter)
have been specially made 
and used to check the resolutions as well as their activities as shown in Fig.~\ref{ULcalibration}.
The $^{241}$Am source hole is covered with 10~$\mu$m Mylar which shifts the energy of the emitted alpha from 5.5~MeV to 4.4~MeV.
When the source was positioned at the center of the detection area, the best resolution of 4.3\% (1~$\sigma$ from a Gaussian fit mean) has been obtained.
However, the resolution degraded to 12\% at the edge of the detection area which is 15~cm away from the center
because the drift electrons are not fully recorded by the anode electrode.
Additionally, we have tested the source on top of the dielectric material to see if there is any difference
due to modified electric field where we found no noticeable effect.
The $^{210}$Po source is created by exposing a low-activity copper plate inside a specially-designed $^{222}$Rn (via the $^{226}$Ra generator) gas container for about 10 days.
After a few days, $^{222}$Rn decays down to $^{210}$Pb which becomes $^{210}$Po on the plate.
With the planar $^{210}$Po source, we were able to reach the best resolution of 3.2\%
at the alpha energy of 5.3~MeV.
The worse resolution from the $^{241}$Am measurement
is due to the imperfect arrangement of the source and the Mylar layer.
Figure~\ref{ULcalibration} shows plots of spectra for each source measurement.

\section{Performance}
\subsection{Alpha identification}
Various groups worldwide with the same detector have been reporting the performance of the detector and their understanding of it~\cite{Abe:2017jzw,McNally:2014eka}.
Triggered events are classified based on the pulse shape analysis
with information from two electrode readouts.
At a humidity level below around 50~ppm$_{v}$, the candidate alpha events from the sample tray can attain
a risetime above 60~$\mu$s. Mid-air events are the alpha decays that happen in the middle of the chamber.
The origin of these mid-air events can be radon contamination from the Argon gas or plumbing pipes, and radon emanation from the detector materials.
Because the mid-air events show up at the detector's lowest sensitivity level and in the region of interest (5--6~MeV), they affect an estimation of the $^{210}$Po alpha activity for ultra-low activity sample measurements.
The ceiling events show a relatively short risetime and the cosmic-ray induced events contains an atypical risetime shape, which can be easily discriminated.
\begin{figure}[!htb]
  \begin{center}
      \includegraphics[width=0.75\textwidth]{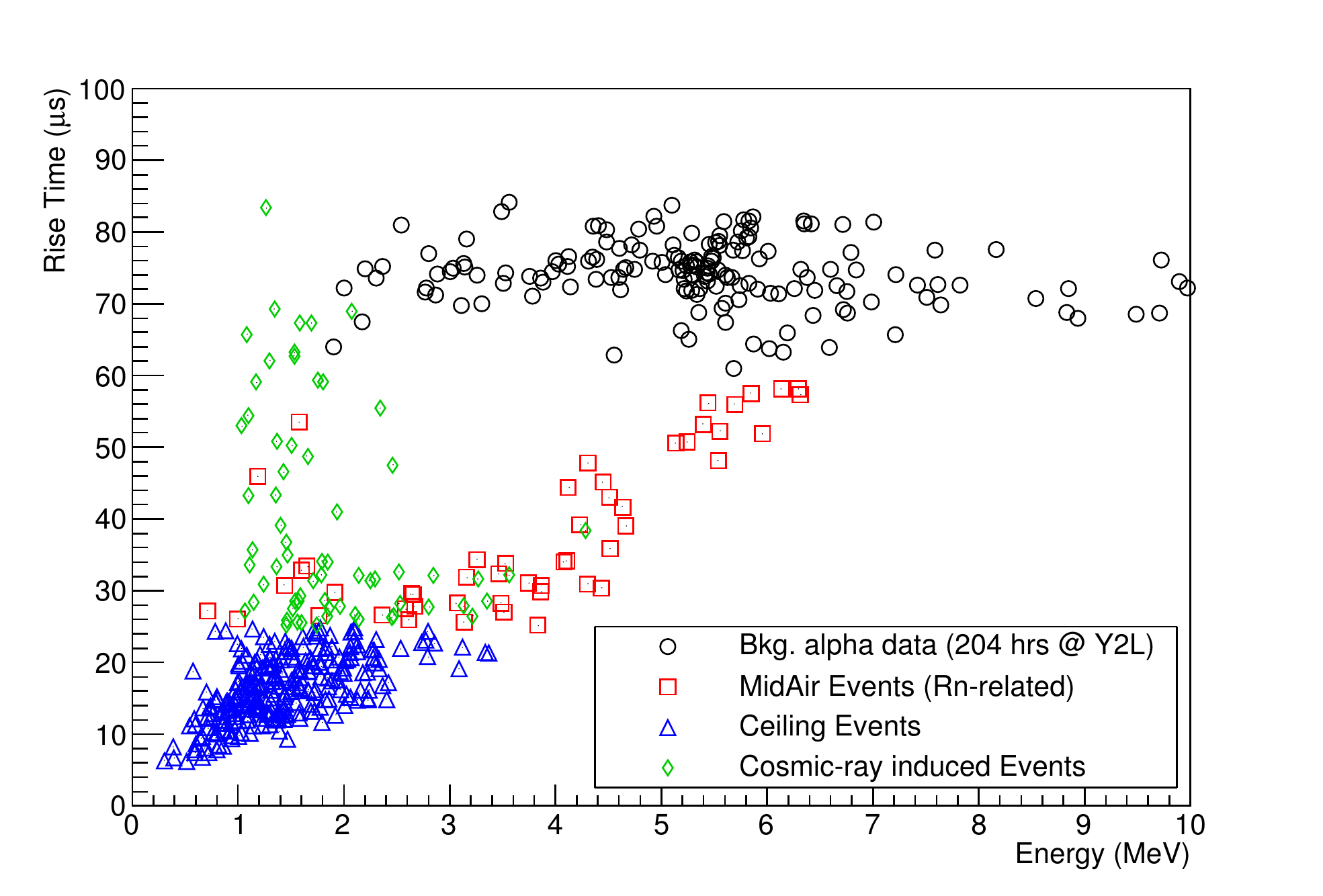}
  \end{center}
  \caption{Alpha particle identification using background data at Y2L.
    Black circles with a risetime above 60~$\mu$s are alpha events.
    Red, blue, and green markers are for the events created from the mid air of the chamber,
   from the ceiling and by cosmic-ray muons, respectively.
  }
  \label{ULpid}
\end{figure}

\subsection{Low-activity lead measurement}
\begin{figure}[!htb]
  \begin{center}
      \includegraphics[width=0.75\textwidth]{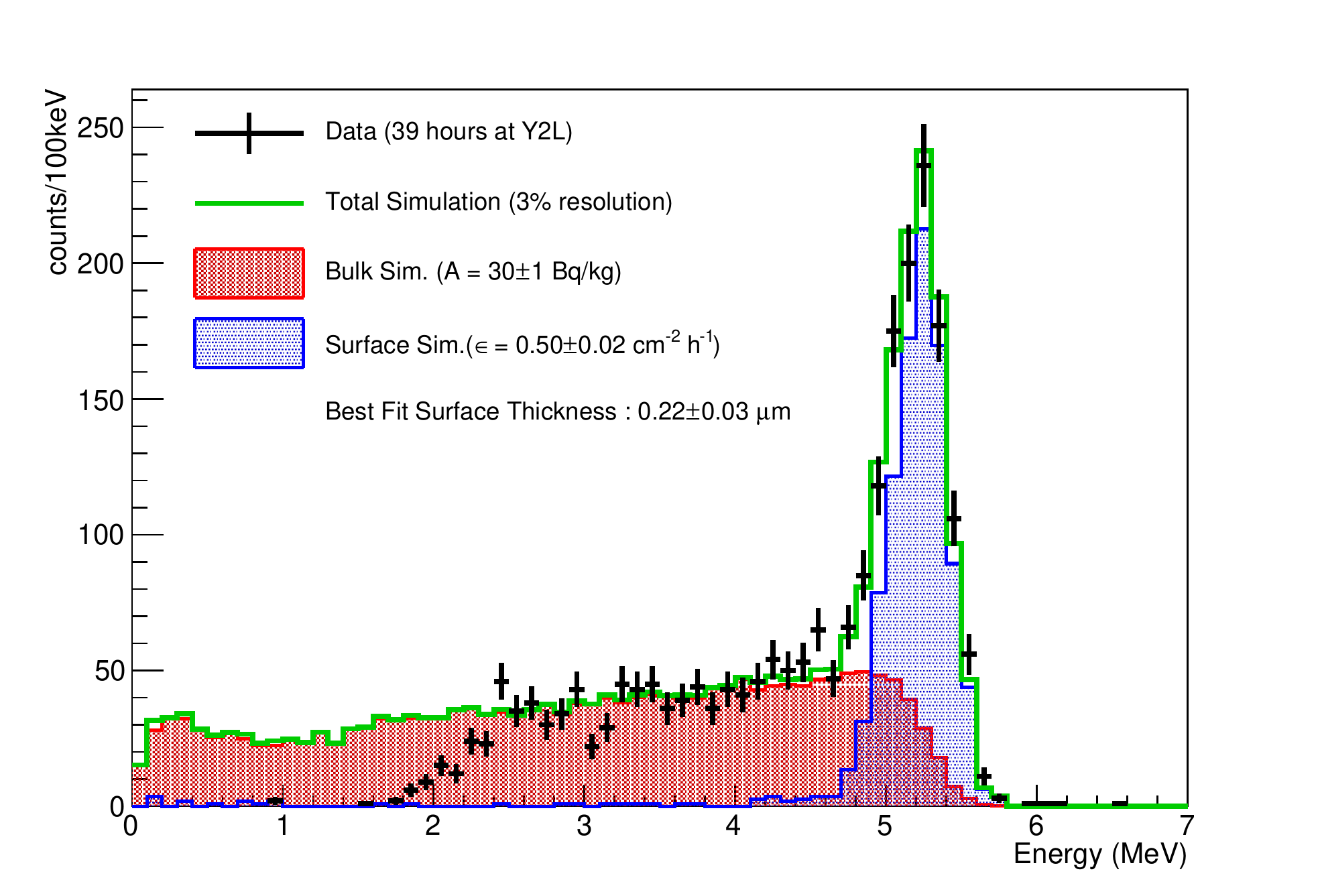}
  \end{center}
  \caption{The lead bar measurement at Y2L.
           The surface alpha activity is obtained separately from the bulk alpha activity by using
           the simulated events with a maximum likelihood fit.
           The blue shaded region is the best fit for $^{210}$Po bulk component while the green shaded region
           is the best fit for $^{210}$Po surface component.
           Note that the fit is performed only above 2.5~MeV region where the efficiency is 100\%.
}
  \label{ULleadbar}
\end{figure}

To confirm detector's sensitivity at the quick measurement level, a lead bar~\footnote{low activity lead from Goslar Inc.}
with a dimension of 10~cm $\times$ 5~cm $\times$ 0.5~cm
has been prepared and counted several times.
The measured spectrum shows a distinct peak at the $^{210}$Po (from $^{210}$Pb decay) alpha energy and a long tail at the left of the peak as shown in Fig.~\ref{ULleadbar}.
The former indicates alpha events mainly originated from the surface of the lead bar while
the latter can be explained as alpha events from the bulk of the lead bar.
We assume all alpha particles observed are emitted only from the $^{210}$Po alpha decay, so the bulk classification arises from a degraded energy due to the self-absorption in a sample.
The bulk $^{210}$Po activities have been measured using the left tail part of a distinctive spectrum in various methods~\cite{Zuzel:2018lrt,Abe:2017jzw}.
\begin{figure}[!htb]
  \begin{center}
      \includegraphics[width=0.75\textwidth]{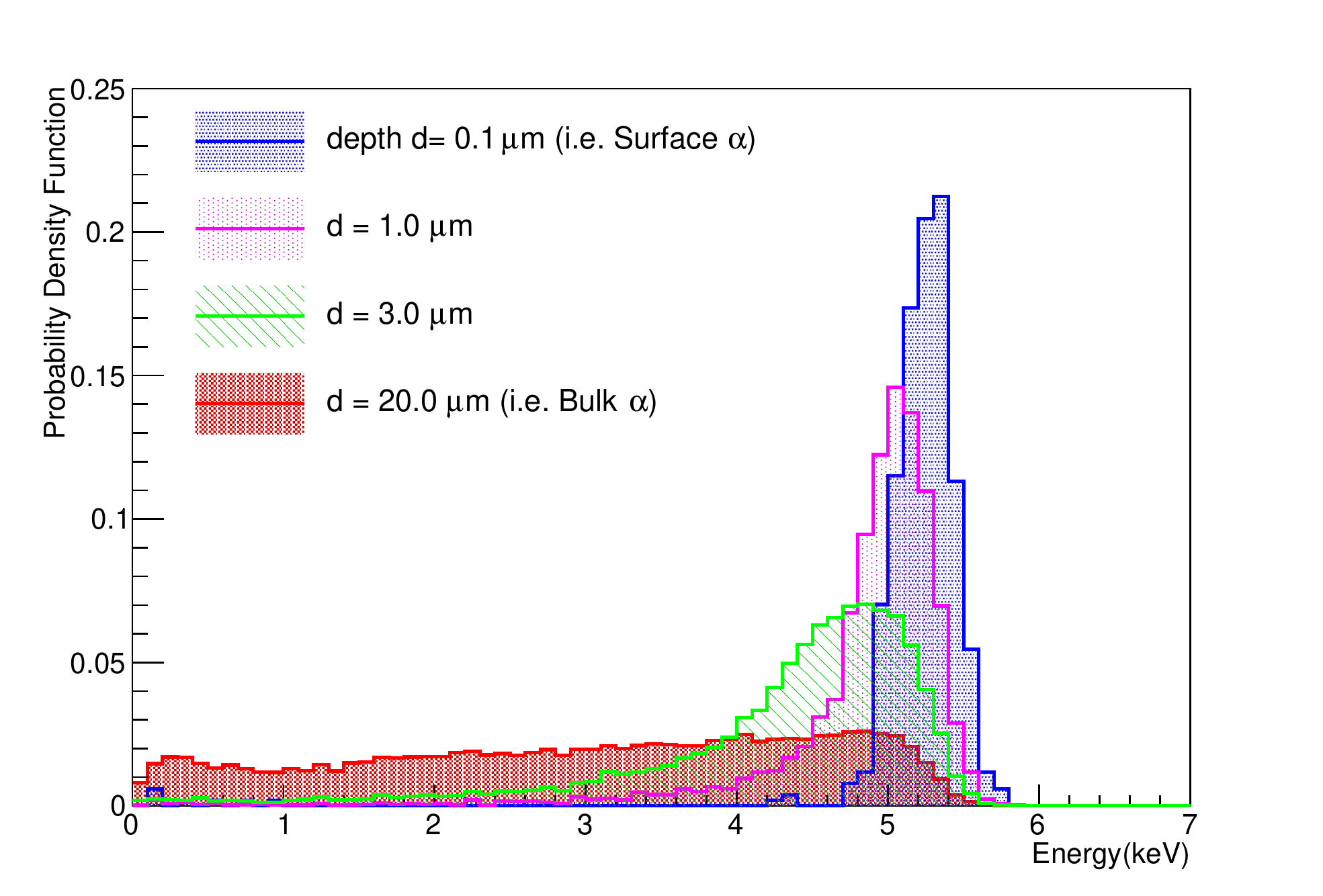}
  \end{center}
  \caption{ Probability density of alpha energy spectra depending on the surface depths
    Alpha events from $^{210}$Po decay are simulated in different depths of the lead sample.
    The thinner depth simulation results in the sharper alpha energy peak as the alpha particle deposits
    energy fully in the detector with no energy loss.
    The energy resolution of 3\% was folded in.
  }
  \label{leadSim}
\end{figure}

Based on this hypothesis, we have performed a maximum likelihood fit using GEANT4 simulation~\cite{geant4} with three parameters including
a normalization of the surface component, a normalization of the bulk component, and the depth (d) of the surface contamination.
The template distribution for the bulk component is obtained by assuming the depth of $^{210}$Po decay position to be smaller than 20~$\mu$m.
We note that the maximum range for the alpha particle to escape from the lead is roughly 16~$\mu$m~\cite{srim}.
The surface template is modeled by changing the depth which is fitted together in the likelihood fitter.
Several example templates are shown in Fig.~\ref{leadSim}.
The energy resolution is assumed to be 3\,\% based on the previous calibrations. 
Parameters' uncertainties are calculated by profiling the likelihood space.
From the best fit, the activities of the bulk component and the surface component are extracted separately.
The bulk alpha activity is fitted to be 0.98$\pm$0.03 cm$^{-2}$~h$^{-1}$ which
converts to 30$\pm$1~Bq/kg.
The best-fit surface activity is 0.50$\pm$0.02 cm$^{-2}$~h$^{-1}$ where
the depth for the surface component is fitted to be 0.22$\pm0.03$~$\mu$m.
The main power for the depth measurement comes from the fact that
the surface component is asymmetric.
The thicker depth shows more asymmetric surface distribution.
This is mainly because of the energy loss of the alpha particle decaying at a certain depth within a sample.
Also, this could be partly explained by the characteristic transport mechanism of $^{210}$Po~\cite{microsegregation}.

The reported value from the producer of the lead bar is 59$\pm$6~Bq/kg which is from measurements of beta particles from $^{210}$Bi decay. Since this value represents the total $^{210}$Pb decay without surface and bulk separation, discrepancy from our measurement is inevitable.
On the other hand, if we combine the bulk and surface components as a single activity, we get 45.5$\pm$2.4~Bq/kg which is consistent at 2.1\,$\sigma$ with the company's measurement. The deviation likely comes from use of different measurement technique, different size of the sample, and independent preparation method.

\begin{figure}[!htb]
  \begin{center}
      \includegraphics[width=0.75\textwidth]{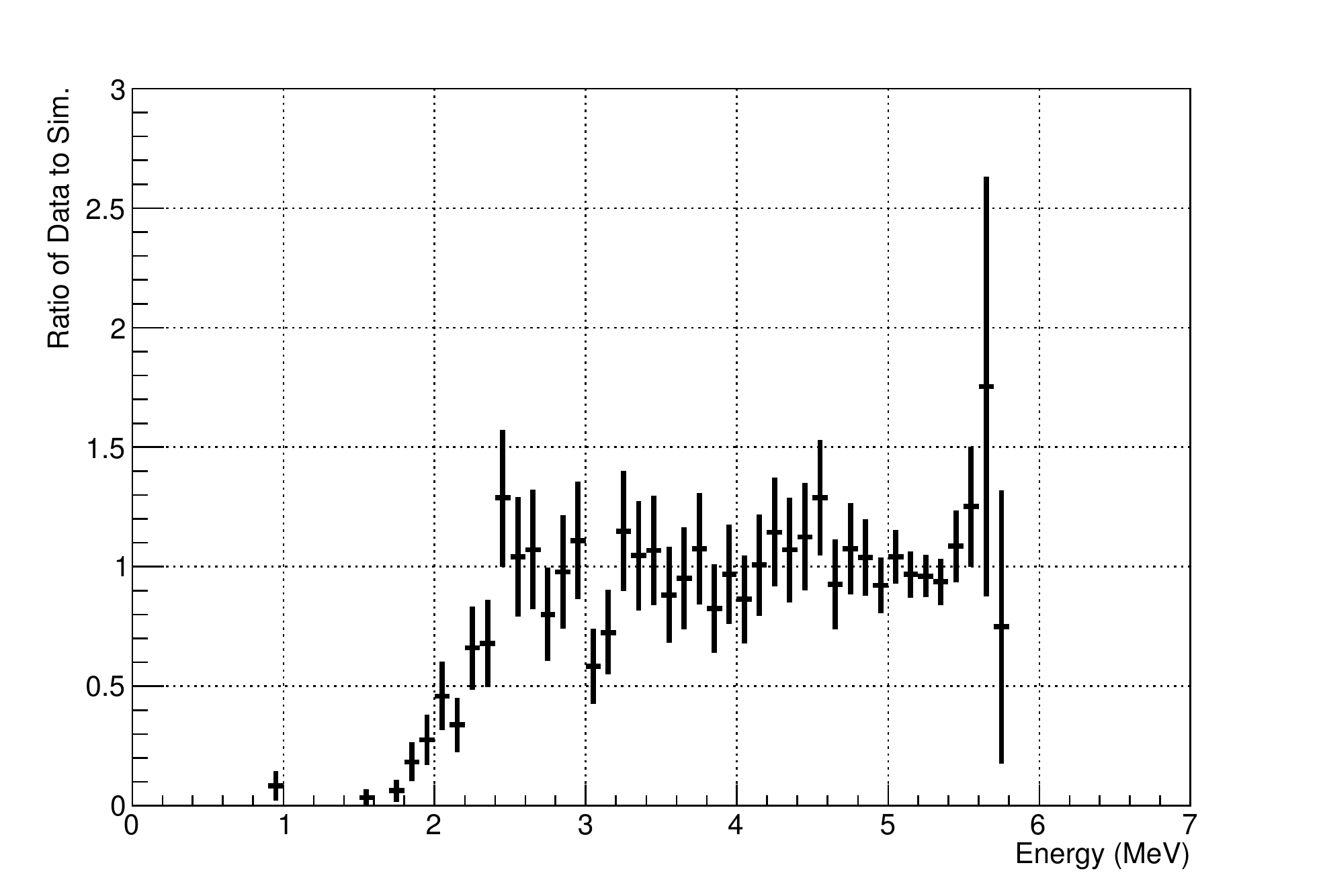}
  \end{center}
  \caption{The efficiency of the detector. By using the lead bar measurement,
    the detector threshold and efficiency have been obtained.
  }   
  \label{ULeff} 
\end{figure}

With the best fit results, the efficiency of the detector
are measured by dividing data events to simulated events.
Figure~\ref{ULeff} shows the energy (trigger) threshold with 50\% efficiency at around 2~MeV.
The energy threshold can be lowered by setting the anode threshold at lower value to collect more lower energy events if necessary.
Later tests show that the energy threshold can be lowered down to around 1~MeV.

\section{Conclusion}
The UltraLo-1800 alpha counter has been running and taking data at 700~m deep Y2L underground facility since 2015.
The detector performs well in the surface alpha measurements of the low radioactivity samples.
With alpha-emitting calibration sources, the alpha counter
was examined in several methods by evaluating resolutions, position-dependence,
and alpha particles from dielectric material.
Additionally, careful tray background measurements allowed us
to identify a new mid-air background events at the lowest activity measurements.
From the $^{210}$Po alpha events in the lead sample,
the surface and bulk components were fit simultaneously with the simulated models, which lead us to understand
the sample contamination depth and the detector's energy threshold.

\section{Acknowledgments}
We thank the Korea Hydro and Nuclear Power (KHNP) Company for providing the underground laboratory space at Yangyang.
We also acknowledge the following supports :
the Institute for Basic Science (IBS) under project code IBS-R016-A1, Republic of Korea.

\end{document}